\documentclass[ALICE,manyauthors]{cernphprep}
\usepackage{hyperref}
\usepackage{lineno}
\usepackage{xcolor}
\usepackage{euscript}
\usepackage{cite}

\usepackage[Symbolsmallscale]{upgreek}
\usepackage{multirow}

\newcommand{\dedx}   {d$E$/d$x$}

\newcommand{\mom}    {\mbox{\rm MeV$\kern-0.15em /\kern-0.12em c$}}
\newcommand{\gmom}   {\mbox{\rm GeV$\kern-0.15em /\kern-0.12em c$}}
\newcommand{\mass}   {\mbox{\rm GeV$\kern-0.15em /\kern-0.12em c^2$}}
\newcommand{\Mmass}  {\mbox{\rm MeV$\kern-0.15em /\kern-0.12em c^2$}}
\newcommand{\pt}     {\ensuremath{p_{\rm T}\xspace}}
\newcommand{\s}      {$\sqrt{s}$}

\newcommand{\kp}     {K$^+$}
\newcommand{\km}     {K$^-$}

\newcommand{\kstar}  {$k^{*}$}
\newcommand{\ap}     {$\overline{\mathrm{p}}$}
\newcommand{\spe}    {$S_\mathrm{T}$}

\newcommand{\opl}    {$\oplus$}
\newcommand{\antik}   {$\mathrm{\overline{K}}\,$} 
\newcommand{\antikz}   {$\mathrm{\overline{K}^0}$}

\begin{document}%

\begin{titlepage}
\PHyear{2019}
\PHnumber{118}      
\PHdate{29 May}  
%

\title{Scattering studies with low-energy kaon-proton femtoscopy in proton--proton collisions at the LHC}
\ShortTitle{Scattering studies with low-energy Kp femtoscopy in pp collisions at the LHC}   

\Collaboration{ALICE Collaboration\thanks{See Appendix~\ref{app:collab} for the list of collaboration members}}
\ShortAuthor{ALICE Collaboration} 

\begin{abstract}
The study of the strength and behaviour of the antikaon-nucleon ($\mathrm{\overline{K}N}$) 
interaction constitutes one of the key focuses of the strangeness sector in low-energy Quantum 
Chromodynamics (QCD). 
In this letter a unique high-precision measurement of the strong interaction between kaons and 
protons, close and above the kinematic threshold, is presented. The femtoscopic measurements of the 
correlation function at low pair-frame relative momentum 
of (\kp p $\oplus$ \km \ap) and (\km p $\oplus$ \kp \ap)
pairs measured in pp collisions at \s\ = 5, 7 and 13 TeV are reported.  A structure observed around 
a relative momentum of 58$~\mom$ in the measured correlation function of 
(\km p $\oplus$ \kp \ap) with a significance of 4.4 $\sigma$ constitutes the first experimental evidence for the opening of the 
$(\mathrm{\overline{K}^0 n} \oplus \mathrm{K^0 \overline{n}})$ isospin breaking 
channel due to the mass difference between charged and neutral kaons.
The measured correlation functions have been compared to J\"ulich and Kyoto models in addition to the Coulomb potential. 
The high-precision data at low relative momenta presented in this work prove femtoscopy to be a 
powerful complementary tool to scattering experiments and provide
new constraints above the $\mathrm{\overline{K}N}$ threshold for low-energy QCD chiral models.

\end{abstract}
\end{titlepage}

\setcounter{page}{2}
The kaon (K) nucleon (N) and anti-kaon (\antik)N interactions constitute the building blocks of low energy QCD with u, d and s quarks, since the effective theories aiming to describe hadron interactions in the non-perturbative energy regime are anchored to these interactions.
Traditionally, the interaction of K and \antik\ with protons and neutrons has been studied by performing scattering experiments at low energies. However, only few measurements
exist and only in a limited energy range~\cite{Humphrey:1962zz, Watson:1963zz, Mast:1975pv, Nowak:1978au, Ciborowski:1982et}.
In such experiments the initial state is fixed, formed by a KN or \antik N pair, and cross-sections of elastic and inelastic final states are measured.\\
These data showed that the K and \antik~behavior with nucleons is very different: while the repulsive nature of \kp p, due to the strong and Coulomb interactions, is well established \cite{Hadjimichef:2002xe}, the strong interacting term of the \km p is instead deeply attractive and characterized by the presence of several coupled-channels, i.e. two-particle systems with energy close to the \km p threshold and carrying the same quantum numbers. 
These coupled-channels contributions are already present in the initial \antik N scattering wave-function and hence influence both the inelastic and the elastic processes~\cite{Shaw:1962zz}.\\
In the \km p system, due to the strangeness $S=-1$ charge of the \antik, already two open coupled-channels appear below threshold: $\Lambda \uppi$ and $\Sigma \uppi$.
Of particular interest is the coupling to the $\Sigma \uppi$ channel since this, along with 
the attractive nature of the \antik  N interaction, leads to the appearance of the $\Lambda(1405)$ resonance just 27~\Mmass~below threshold. Indeed, this resonance is interpreted as the only  $\Sigma\uppi$-$\mathrm{\overline{K}N}$ molecular state~\cite{Dalitz:1959dn,Dalitz:1960du,Hall:2014uca}.
The available theoretical approaches ~\cite{KAISER1995325,OSET199899,OLLER2001263,LUTZ2002193,Hyodo:2007jq,KAMIYA201641,Revai:2017isg,Mai:2014xna,Borasoy:2006sr,Cieply:2015pwa} are constrained above the \antik N threshold, but since the experimental data are scarce, these constraints are rather loose resulting in rather significant differences below threshold. 
Experimental constraints on the $\mathrm{\overline{K}N}$ interaction and on the interplay between both $\mathrm{\overline{K}N}$ and $\Sigma \uppi$ poles, are fundamental to reproduce the properties of the $\Lambda(1405)$~\cite{Hemingway:1984pz,Braun:1977wd,Thomas:1973uh,Agakishiev:2012xk,Moriya:2013eb}.\\
Approximately 5 MeV above threshold, the \antikz n channel opens up due to the breaking of the isospin symmetry.  
The \antik n-KN coupling is also very important to understand the interaction and structure of the $\Lambda$(1405) and its 
effect should be visible in the total \km p cross-section measured in scattering experiments as a clear cusp-like structure for a kaon incident momentum of  $p_{\mathrm{lab}}=89~$\mom~~\cite{Ikeda:2012au}.  However, this peak has not been experimentally observed yet due to the large uncertainties of the  data~\cite{Mast:1975pv,Ciborowski:1982et,Sakitt:1965kh}.\\
In order to constrain the contributions of the coupled-channels and to provide a complete description of the \antik N interaction, precise data close to threshold are needed and effects of coupled-channels lying close to threshold must be explicitly taken into account in any process between a \antik\  and a nucleon.\\
The  measurement  of  kaonic  hydrogen~\cite{Bazzi:2011zj},  which nowadays constitutes  the  most  precise constraint at threshold, and the obtained results on the \antik N scattering parameters include the coupled-channel contributions only in an effective way.

Recently, the femtoscopy technique \cite{Lednicky:2005af,Lisa:2005dd}, which measures the correlation of particle pairs at low relative momentum,
has provided high 
precision data on different baryon--baryon pairs~\cite{Kisiel:2014mma,Adamczyk:2015hza,Acharya:2018gyz}, indicating a great sensitivity to the underlying strong potential. Contrary to the scattering, in femtoscopy 
only the final state is measured and different initial states are allowed. In the \km p system, this translates into an extreme sensitivity of the correlation function to the introduction of the different coupled-channels, which affect both shape and magnitude of the femtoscopic signal~\cite{Haidenbauer:2018jvl}.

The femtoscopic measurement of Kp pairs ((\kp p \opl\ \km \ap)
and (\km p \opl\ \kp \ap)) from pp collisions at different energies presented in this Letter shows experimentally
for the first time the impact of coupled-channels effect on the momentum correlation function. Comparison with recent models including or partially including coupled-channel contributions are presented.
The same-charge pairs (\kp p \opl\ \km \ap), because of the well described interaction 
and the lack of coupled-channel effects, are used as a benchmark to test the sensitivity of 
the correlation function to the strong interaction.\\
The analysis presented here is based on minimum bias triggered pp collisions collected by the
ALICE experiment~\cite{Abelev:2014ffa} at the LHC in 2010, 2015, 2016 and 2017 at three 
different collision energies (\s~=~5~TeV, 7 TeV, and 13 TeV). 
The correlation function $C$(\kstar) is measured as a function of the momentum difference of the pair \kstar~=~$\frac{1}{2}(\vec{p_1}^* - \vec{p_2}^*)$, where $\vec{p_1}^*$ and $\vec{p_2}^*$ 
are the momenta of the two particles in the pair rest frame. It is 
defined as $C$(\kstar) = $\EuScript{N}$ $A$(\kstar)/$B$(\kstar), where $A$(\kstar) is 
the measured distribution of pairs from the same event, $B$(\kstar) is the reference distribution 
of pairs from mixed events and $\EuScript{N}$ is a normalization parameter. 
The denominator, $B$(\kstar), is formed by mixing particles from one event with particles from a  
pool of other events with comparable number of charged particles at mid-rapidity~\cite{ALICE-PUBLIC-2019-006} and comparable interval of the collision primary vertex coordinate along the beam axis, $V_z$ interval (\mbox{$\Delta V_z$ = 2~cm}).
The normalization parameter $\EuScript{N}$ is chosen such that the mean value of the correlation  
function equals unity for 400~$<$~\kstar~$<$~600~\mom.\\
The main sub-detectors used in this analysis are: the V0 detectors~\cite{Abbas:2013VZERO}, 
which are used as trigger detectors, the Inner Tracking System (ITS)~\cite{Aamodt:2010aa}, 
the Time Projection Chamber (TPC) \cite{Alme:2010TPC} and the Time-of-Flight (TOF) detector
\cite{Akindinov:2013tea}. The ITS, TPC and TOF are located inside a 0.5~T solenoidal magnetic 
field and are used to track and identify charged particles. 
In order to ensure a uniform acceptance at mid-rapidity, events were 
selected by requiring the $V_z$ of the event to be within 10 cm from the center of the ALICE 
detector. 
The rejection of pile-up is performed by exploiting the innermost silicon detector (SPD, part of ITS) vertexing capabilities, following the same procedure described in~\cite{Acharya:2018gyz, Acharya:2019yvb}.
After the application of the event selection criteria, about 874~million, 374~million, and 1~billion minimum bias pp events were analyzed at \s~=~5~TeV, 7~TeV, and 13~TeV, respectively. \\
As recently proposed in~\cite{Acharya:2019idg}, in order to reduce 
the contribution from the mini-jet background in pp collisions, the events 
were classified according to their transverse sphericity (\spe), an observable which is known to be correlated with the number of hard 
parton--parton interactions in each event~\cite{Abelev:2012sk}. 
An event with only one hard parton--parton interaction will generally 
produce a jet-like distribution that yields low sphericity, while an event 
with several independent hard parton--parton interactions can yield higher 
sphericity. 
To reduce the strong mini-jet background at low momenta, only events 
with \spe, defined as in~\cite{Acharya:2019idg}, larger than 0.7 were considered in this analysis. \\
Charged particles were tracked and selected using the same criteria described in~\cite{Acharya:2018gyz}. 
The charged kaons and protons were identified in a wide transverse momentum (\pt) interval 
(0.15~$<$~\pt~$<$~1.4~\gmom\ for kaons and 0.4~$<$~\pt~$<$~3~\gmom\ for protons) using 
the information provided by the TPC and the TOF detectors. 
The deviation of the measured specific ionization energy loss (\dedx) in the TPC from 
the Bethe-Bloch parametrization was required to be within three 
standard deviations ($\sigma_{\mathrm{TPC}}$). 
For kaons with \pt~$>$~0.4~\gmom\ and protons with \pt~$>$~0.8~\gmom, a similar method was applied 
for the particle identification using the TOF, where, on top of TPC selection, a selection based on 
a maximum three standard deviation difference from the expected signal at a given momentum was applied. 
Tracks identified ambiguously as belonging to both a proton and a kaon, were discarded. 
In order to remove the large fraction of e$^+$e$^-$ pairs that can affect the 
extraction of the correlation function of the opposite-charge pairs, a selection on 
the \pt\ of kaon and protons was applied: kaon candidates are excluded if 
\mbox{0.3~$<$\pt~$<$~0.4~\gmom}, while proton candidates 
are excluded in the interval between 0.6~$<$~\pt~$<$~0.8~\gmom. 
The purity of the selected particle samples, determined by Monte Carlo 
simulations, is larger than 99\% in the considered \pt\ intervals for all the 
analyzed dataset. 
The systematic uncertainties of the measured $C(k^*)$ were evaluated for each 
\kstar\ interval by varying event and track selection criteria. 
The event sample was varied by changing the selection on the $V_z$ position from $\pm$~10~cm 
to $\pm$~7~cm and by varying the sphericity 
of the accepted events from \spe~$>$~0.7 to  \spe~$>$~0.6 and \spe~$>$~0.8.  
Systematic uncertainties related to the track selection criteria were studied by varying 
the selection on the  Distance of Closest Approach  in the transverse plane direction 
within the experimental resolution.    
To study systematic effects related to particle identification, the number of standard 
deviations around the energy loss expected for kaons and protons in the TPC and, similarly, for 
the time-of-flight in the TOF was modified from 3$\sigma$ to 2$\sigma$. 
For each source, the systematic uncertainty was estimated as the root-mean-square (RMS) 
of the deviations. The total systematic uncertainty was calculated as the quadratic sum of 
each source's contribution and amounts to about 3\% in the considered \kstar\ intervals.

\begin{figure}[!htb]  
\begin{center}
\includegraphics[width=1\textwidth]{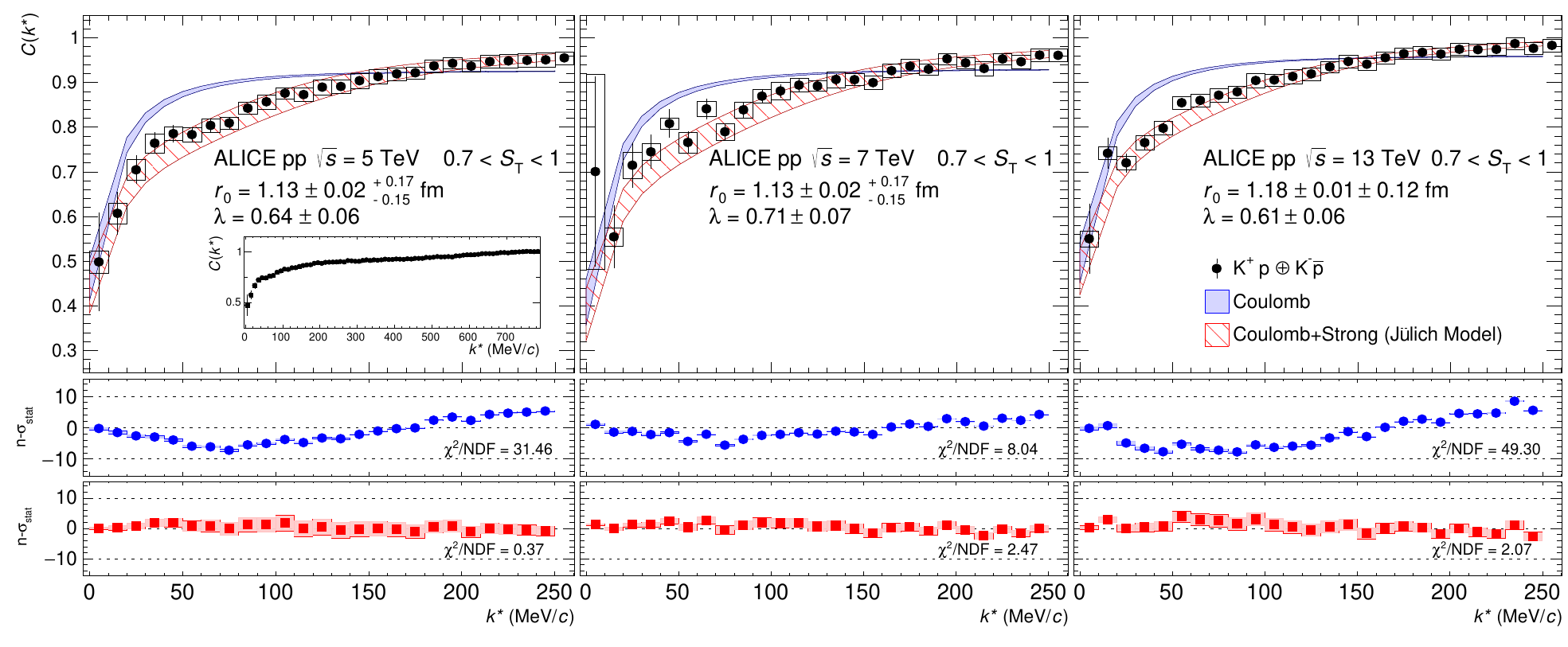}
\end{center}
\caption{(\kp p \opl\ \km \ap) correlation functions obtained from pp collisions at 
\s~=~5~TeV (left), 7 TeV (middle) and 13 TeV (right). The inset shows the correlation function evaluated for pp collisions at 
\s~=~5~TeV  in a wider \kstar interval. The measurement is shown by the black markers, the vertical lines and the boxes represent the statistical and systematic uncertainties respectively. Bottom panels represent comparison with models as described in the text.}
\label{fig:KppCFfit}
\end{figure}

\begin{figure}[!htb]  
\begin{center}
\includegraphics[width=1\textwidth]{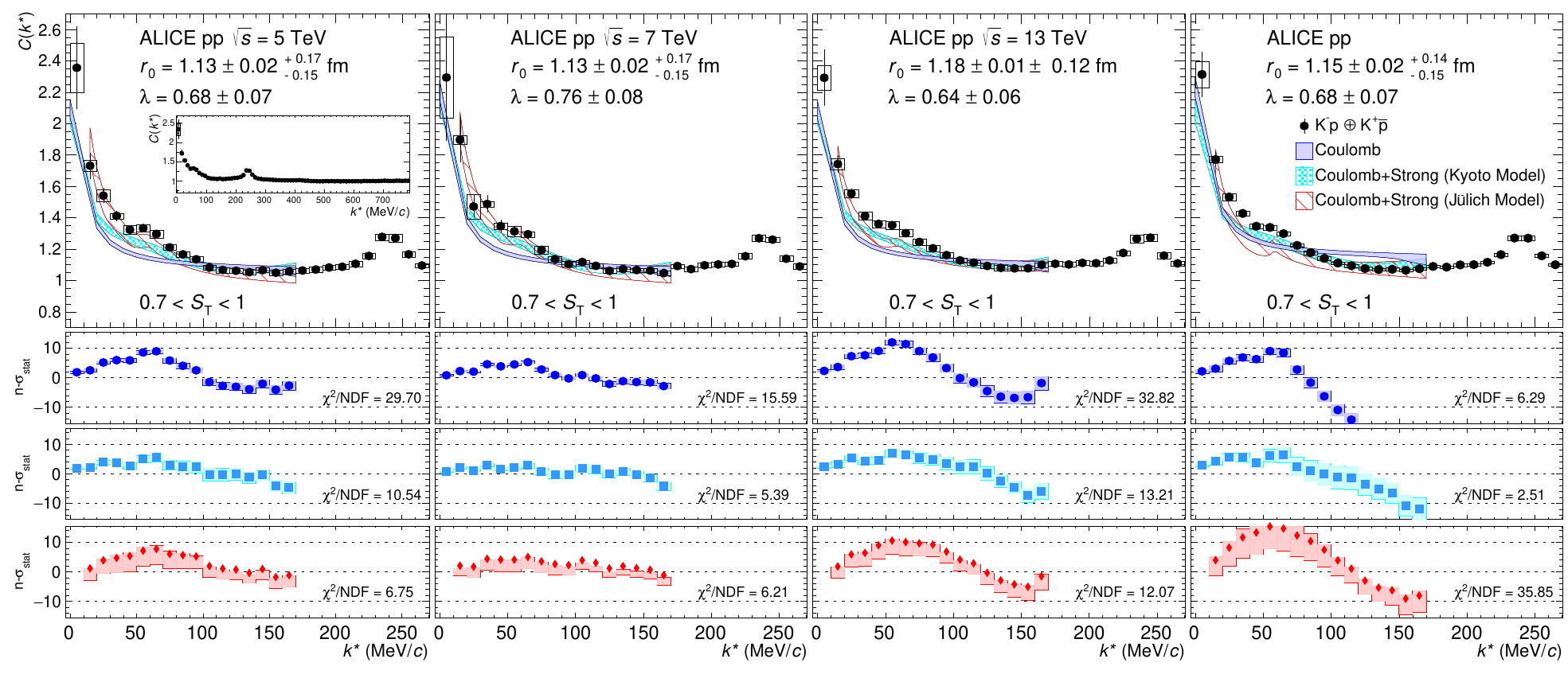}
\end{center}
\caption{(\km p \opl\ \kp \ap) correlation functions obtained (from left to right) from pp collisions at 
\s~=~5~TeV, 7 TeV, 13 TeV. The fourth panel shows the combined results at the three colliding energies, 
The number of pairs  in each data sample as been used as weight.  The inset shows the correlation function evaluated for pp collisions at 
\s~=~5~TeV  in a wider \kstar interval. The measurement is presented by the black markers, the vertical lines and the boxes represent the statistical and systematic uncertainties respectively. Bottom panels represent comparison with models as described in the text.}
\label{fig:KmpCFfit}
\end{figure}

The measured correlation functions for (\kp p \opl\ \km \ap) and (\km p \opl\ \kp \ap) 
are shown in the upper panels of Fig.~\ref{fig:KppCFfit} and Fig.~\ref{fig:KmpCFfit}. In both figures, each panel corresponds to a different collision energy, as indicated in the legend.
The structure that can be seen in the (\km p \opl\ \kp \ap) correlation function at 
$k^*$ around 240~\mom\ in Fig.~\ref{fig:KmpCFfit} is consistent with the $\Lambda$(1520) which decays into $\mathrm{{K}^- p}$, with a center-of-mass momentum for the particle pair of 243~\mom~\cite{Tanabashi:2018oca}. 
The correlation function of (\km p \opl\ \kp \ap) exhibits also a clear structure between 
50 and 60 \mom\ for the 
three collision energies. The \kstar\ position of the structure is consistent with the threshold of the $\mathrm{\overline{K}^0n}$ ($\mathrm{{K}^0\overline{n}}$) channel opening at $p_{\mathrm{lab}}$~=~89~\mom~\cite{Mast:1975pv,Ciborowski:1982et,Sakitt:1965kh} which corresponds to \kstar~=~58~\mom.  
In order to quantify the significance of this structure, and since the three measured 
distributions are mutually compatible, the $C(k^*)$ measured at the three different 
energies were summed using the number of pairs  in each data sample as a weight. 
The resulting $C(k^*)$ was interpolated with a spline considering the 
statistical uncertainties and the derivative of the spline was then evaluated~\cite{ALICE-PUBLIC-2019-006}. 
A change in the slope of the derivative consistent with a cusp effect in the \kstar\ region 
between 50 and 60~\mom\ at the level of 4.4$\sigma$ has been observed, to be compared 
with a significance of 30$\sigma$ for $\Lambda$(1520). 
The measurement presented here is therefore the first experimental evidence for the opening of the $\mathrm{\overline{K}^0n}$ ($\mathrm{{K}^0 \overline{n}}$) channel,  showing that the femtoscopy technique is a unique tool to study the $\mathrm{\overline{K}p}$ interaction and coupled-channel effects.

The experimental correlation functions were also used to test different potentials 
to describe the interaction between \kp p (\km \ap) and \km p (\kp \ap).  
The measured correlation function $C(k^*)$ is compared with a theoretical function 
using the following equation

\begin{equation}
C(k^*) = (a + b \cdot k^*) \cdot \left[ 1+\lambda \cdot (C(k^*)^{theoretical} - 1)\right], 
\label{eq:cf1}
\end{equation}

where the baseline $(a + b \cdot k^*)$ is introduced to take into account the 
remaining  non-femtoscopic background contributions related to momentum-energy conservation which might be present also 
after the \spe\ selection. 
The slope, $b$, of the baseline is fixed from Monte Carlo simulations based on
PYTHIA~6~\cite{Sjostrand:2006za} and PYTHIA~8~\cite{Sjostrand:2007gs}, while 
the normalization, $a$, 
is a free parameter. In order to assign a systematic uncertainty related to the slope of the baseline, the $b$ parameter has been varied by its uncertainty as obtained from the Monte Carlo simulation ($\pm$ 10\%) and the fit repeated. 
The parameter $\lambda$ represents the fraction of primary pairs in the analyzed sample 
multiplied by the purity of the same sample and is fixed by fitting Monte Carlo (MC) 
templates to the experimental distributions of DCA$_{xy}$ of kaons and protons, 
similarly to what is described in~\cite{Acharya:2018gyz}. 

The model correlation function, $C(k^*)^{theoretical}$, is evaluated using the 
CATS framework~\cite{Mihaylov:2018rva}. 
The $\lambda$ parameters obtained for each analyzed dataset are reported in each panel 
of Fig.~\ref{fig:KppCFfit} and Fig.~\ref{fig:KmpCFfit} for same-charge and opposite-charge 
Kp pairs, and vary from 0.61 to 0.76 for each considered set. 
A systematic uncertainty of $\pm$10\% is associated with the $\lambda$ parameters. 
This uncertainty was estimated by varying the Monte Carlo templates used in the 
feed-down estimation procedure based on PYTHIA~6~\cite{Sjostrand:2006za} for the analysis 
at \s~=~7~TeV and based on PYTHIA~8~\cite{Sjostrand:2007gs} for the 
analyses performed at \s~=~5~TeV and 13~TeV, and varying the transport code used in 
the simulation from GEANT3~\cite{Brun:1994} to GEANT4~\cite{Agostinelli:2002hh}. \\
The effects related to momentum resolution effects are accounted for
by correcting the theoretical correlation function, similarly to what  shown in~\cite{Acharya:2018gyz} and~\cite{Acharya:2019yvb}.
The theoretical correlation function $C(k^*)^{theoretical}$ depends not 
only on the interaction between particles, but also on the profile and the size of 
the particle emitting source. Under the assumption that there is a common Gaussian source 
for all particle pairs produced 
in pp collisions at a fixed energy, the size of the source considered in the present analysis is 
fixed from the baryon--baryon analyses described in~\cite{Acharya:2018gyz} and \cite{Acharya:2019yvb}. 
The impact of strongly decaying resonances (mainly K$^*$ decaying into K and $\Delta$ decaying into p) on the determination of the radius for Kp pairs was studied
using different Monte Carlo simulations \cite{Sjostrand:2006za, Sjostrand:2007gs} and found to be 10\%. 
This contribution was linearly added to the systematic uncertainty associated with the radius. 
The radii of the considered Gaussian sources are $r_0 = 1.13 {\pm 0.02^{+0.17}_{-0.15}}$~fm~\cite{Acharya:2018gyz} 
for collisions at \s\ = 5 and 7 TeV, and $r_0 = 1.18 \pm 0.01 \pm 0.12 $~fm~\cite{Acharya:2019yvb} 
for the \s\ = 13 TeV collisions. 

The comparison of the measured $C(k^*)$ for same-charge Kp 
pairs with different models is shown in Fig.~\ref{fig:KppCFfit}. 
Each panel presents the results at different collision 
energy and the comparison with two different scenarios.
The blue band represents the correlation 
function evaluated as described in Eq.~(\ref{eq:cf1}), assuming only the 
presence of the Coulomb potential to evaluate the $C(k^*)^{theoretical}$ term. 
The red band represents the correlation function assuming 
the strong potential implemented in the J\"ulich model \cite{Haidenbauer:2010ch} 
in addition to the Coulomb potential. The latter has been implemented using the Gamow factor~\cite{Adam:2015vja}.
In the bottom panels, the difference between data and 
model evaluated in the middle of each \kstar\ interval, and divided by statistical error of data for the three considered collision energies are shown. The width of the bands
represents the n-$\sigma$ range associated to the model variations. The reduced $\chi^2$ are also shown.
This comparison reveals that the Coulomb interaction is not able to 
describe the data points, as expected, while the introduction of a strong potential allows to reproduce 
consistently the data when the same source radius as for baryon-baryon pairs is considered. Hence, the measured correlation functions are sensitive to the
strong interaction and can be used to test different strong potentials for the \km p system, assuming a common source for all the Kp pairs produced in a collision. 

Similar to Fig.~\ref{fig:KppCFfit} for like-sign pairs, Fig.~\ref{fig:KmpCFfit} shows the data-model comparison for unlike-sign pairs.
The measured $C(k^*)$ is reported for the three different collision energies and the $C(k^*)$ distributions 
were compared with different interaction models. 
Since all the models considered in this letter do not take the presence of $\Lambda$(1520) into account, only the region below 170~\mom\ is considered in the comparison.   
The blue bands show results obtained using CATS with a Coulomb potential only.\\
The remaining curves include, on top of the Coulomb attraction, different 
descriptions of the $\mathrm{\overline{K}N}$ strong interaction. The width of each band accounts for the uncertainties in the $\lambda$ parameters, the source radius and the baseline. The light blue bands corresponds to the Kyoto 
model calculations with approximate boundary conditions on the \km p wave-function which neglect the contributions from  $\mathrm{\Sigma\uppi}$ and $\mathrm{\Lambda\uppi}$ coupled-channels~\cite{Miyahara:2015bya,Ohnishi:2017uni,Cho:2017dcy,Ikeda:2011pi,Ikeda:2012au}. Moreover, this version of the Kyoto model is performed in the so-called isospin basis and hence does not include the mass difference between \km\ and 
$\overline{\mathrm{K}}^0$: no  cusp-like structure are foreseen by the model in $C(k^*)$.
\\
The introduction of coupled-channel contributions in the correlation function has been shown to result in additional attractive terms enhancing the signal, in particular in the low \kstar\ region~\cite{Haidenbauer:2018jvl}. As expected, the Kyoto results clearly underestimate the data at low momenta where the $\mathrm{\Sigma\uppi}$ channel is particularly relevant.\\
The red bands indicate results obtained with the  J\"ulich strong potential, recently updated to reproduce the kaonic atom results from SIDDHARTA collaboration~\cite{Haidenbauer:2018jvl}. This model includes explicitly both $\mathrm{\Sigma\uppi}$ and $\mathrm{\Lambda\uppi}$ coupled-channels below threshold and the \km-- $\overline{\mathrm{K}}^0$ mass difference, reflected in the presence of a cusp structure. Accordingly, the comparison with data shows a better agreement with respect to the Kyoto model, but the region of \kstar\ below 100~\mom\ is nevertheless not fully reproduced and the shape of the correlation function 
deviate from the data around the cusp.\\
The overall tension between data and the models is not surprising since the latter were fitted to only reproduce scattering data above threshold (providing constraints for \kstar $\geq 70$ \mom) 
and the SIDDHARTA results at threshold~\cite{Bazzi:2011zj}. 

To test the stability of the results, the measured $C(k^*$) without any \spe\ cut was used and the background from mini-jets and other kinetically correlated pairs was subtracted by using a Monte Carlo simulation based on PYTHIA~8~\cite{Sjostrand:2007gs}, using a procedure similar to the one described in~\cite{Acharya:2018kpo}.
Applying this method the comparison between data and models is consistent within statistical uncertainties with the one obtained using the sphericity selection.

To summarize, the momentum dependent correlations of same-charge and opposite-charge Kp pairs ((\kp p \opl\ \km \ap) 
and (\km p \opl\ \kp \ap)) were measured using the two-particle correlation function in pp collisions at different collision energies. 
A structure around \kstar~=~58 \mom\ in the measured correlation function of 
(\km p \opl\ \kp \ap) was observed. 
The significance of such a structure was evaluated by combining the results from the three analyzed datasets and by interpolating the total correlation function with a spline. 
By studying the variation in the slope of the derivative of such spline in the range $50~\leq k^\ast \leq 60 $~\mom, the kinematic cusp was assessed at a $4.4\sigma$ level.
The observed structure is consistent with the opening of the 
$\mathrm{\overline{K}^0n}$ channel ($p_{\mathrm{lab}} \sim $~89~\mom). This measurement 
represents the first experimental evidence for 
the $\mathrm{\overline{K}^0n}$ ($\mathrm{{K}^0\overline n}$) isospin breaking coupled-channel and shows experimentally the effect of coupled-channel contributions on the correlation function. \\
The measured $C({k^*})$ were compared to different interaction scenarios.
The (\kp p \opl\ \km \ap) correlation functions were proven to be 
sensitive to the strong interaction, since a Coulomb-only hypothesis is 
insufficient to describe the data. The inclusion of the strong interaction 
via the J\"ulich model results in a reasonable description of the data within uncertainties. 
The (\km p \opl\ \kp \ap) 
correlation functions at low \kstar\ cannot be fully reproduced by the 
considered potentials. Nevertheless, model including explicitly coupled-channel contributions shows a better agreement with data.
The data presented here represent the most precise experimental information for the KN interaction and provide new 
constraints for future low-energy phenomenological QCD calculations can be used 
to shed light on the nature of the $\overline{\mathrm{K}}$N interaction. 


\newenvironment{acknowledgement}{\relax}{\relax}
\begin{acknowledgement}
\section*{Acknowledgements}
The ALICE Collaboration is grateful to Prof.~Tetsuo Hyodo and Prof.~Johann Haidenbauer
for the valuable suggestions and discussions.

The ALICE Collaboration would like to thank all its engineers and technicians for their invaluable contributions to the construction of the experiment and the CERN accelerator teams for the outstanding performance of the LHC complex.
The ALICE Collaboration gratefully acknowledges the resources and support provided by all Grid centres and the Worldwide LHC Computing Grid (WLCG) collaboration.
The ALICE Collaboration acknowledges the following funding agencies for their support in building and running the ALICE detector:
A. I. Alikhanyan National Science Laboratory (Yerevan Physics Institute) Foundation (ANSL), State Committee of Science and World Federation of Scientists (WFS), Armenia;
Austrian Academy of Sciences, Austrian Science Fund (FWF): [M 2467-N36] and Nationalstiftung f\"{u}r Forschung, Technologie und Entwicklung, Austria;
Ministry of Communications and High Technologies, National Nuclear Research Center, Azerbaijan;
Conselho Nacional de Desenvolvimento Cient\'{\i}fico e Tecnol\'{o}gico (CNPq), Universidade Federal do Rio Grande do Sul (UFRGS), Financiadora de Estudos e Projetos (Finep) and Funda\c{c}\~{a}o de Amparo \`{a} Pesquisa do Estado de S\~{a}o Paulo (FAPESP), Brazil;
Ministry of Science \& Technology of China (MSTC), National Natural Science Foundation of China (NSFC) and Ministry of Education of China (MOEC) , China;
Croatian Science Foundation and Ministry of Science and Education, Croatia;
Centro de Aplicaciones Tecnol\'{o}gicas y Desarrollo Nuclear (CEADEN), Cubaenerg\'{\i}a, Cuba;
Ministry of Education, Youth and Sports of the Czech Republic, Czech Republic;
The Danish Council for Independent Research | Natural Sciences, the Carlsberg Foundation and Danish National Research Foundation (DNRF), Denmark;
Helsinki Institute of Physics (HIP), Finland;
Commissariat \`{a} l'Energie Atomique (CEA), Institut National de Physique Nucl\'{e}aire et de Physique des Particules (IN2P3) and Centre National de la Recherche Scientifique (CNRS) and R\'{e}gion des  Pays de la Loire, France;
Bundesministerium f\"{u}r Bildung und Forschung (BMBF) and GSI Helmholtzzentrum f\"{u}r Schwerionenforschung GmbH, Germany;
General Secretariat for Research and Technology, Ministry of Education, Research and Religions, Greece;
National Research, Development and Innovation Office, Hungary;
Department of Atomic Energy Government of India (DAE), Department of Science and Technology, Government of India (DST), University Grants Commission, Government of India (UGC) and Council of Scientific and Industrial Research (CSIR), India;
Indonesian Institute of Science, Indonesia;
Centro Fermi - Museo Storico della Fisica e Centro Studi e Ricerche Enrico Fermi and Istituto Nazionale di Fisica Nucleare (INFN), Italy;
Institute for Innovative Science and Technology , Nagasaki Institute of Applied Science (IIST), Japan Society for the Promotion of Science (JSPS) KAKENHI and Japanese Ministry of Education, Culture, Sports, Science and Technology (MEXT), Japan;
Consejo Nacional de Ciencia (CONACYT) y Tecnolog\'{i}a, through Fondo de Cooperaci\'{o}n Internacional en Ciencia y Tecnolog\'{i}a (FONCICYT) and Direcci\'{o}n General de Asuntos del Personal Academico (DGAPA), Mexico;
Nederlandse Organisatie voor Wetenschappelijk Onderzoek (NWO), Netherlands;
The Research Council of Norway, Norway;
Commission on Science and Technology for Sustainable Development in the South (COMSATS), Pakistan;
Pontificia Universidad Cat\'{o}lica del Per\'{u}, Peru;
Ministry of Science and Higher Education and National Science Centre, Poland;
Korea Institute of Science and Technology Information and National Research Foundation of Korea (NRF), Republic of Korea;
Ministry of Education and Scientific Research, Institute of Atomic Physics and Ministry of Research and Innovation and Institute of Atomic Physics, Romania;
Joint Institute for Nuclear Research (JINR), Ministry of Education and Science of the Russian Federation, National Research Centre Kurchatov Institute, Russian Science Foundation and Russian Foundation for Basic Research, Russia;
Ministry of Education, Science, Research and Sport of the Slovak Republic, Slovakia;
National Research Foundation of South Africa, South Africa;
Swedish Research Council (VR) and Knut \& Alice Wallenberg Foundation (KAW), Sweden;
European Organization for Nuclear Research, Switzerland;
National Science and Technology Development Agency (NSDTA), Suranaree University of Technology (SUT) and Office of the Higher Education Commission under NRU project of Thailand, Thailand;
Turkish Atomic Energy Agency (TAEK), Turkey;
National Academy of  Sciences of Ukraine, Ukraine;
Science and Technology Facilities Council (STFC), United Kingdom;
National Science Foundation of the United States of America (NSF) and United States Department of Energy, Office of Nuclear Physics (DOE NP), United States of America.
\end{acknowledgement}


%
%

\bibliographystyle{utphys}
\bibliography{bibliog}

\newpage
\appendix

\section{The ALICE Collaboration}
\label{app:collab}

\begingroup
\small
\begin{flushleft}
S.~Acharya\Irefn{org141}\And 
D.~Adamov\'{a}\Irefn{org93}\And 
S.P.~Adhya\Irefn{org141}\And 
A.~Adler\Irefn{org74}\And 
J.~Adolfsson\Irefn{org80}\And 
M.M.~Aggarwal\Irefn{org98}\And 
G.~Aglieri Rinella\Irefn{org34}\And 
M.~Agnello\Irefn{org31}\And 
N.~Agrawal\Irefn{org10}\And 
Z.~Ahammed\Irefn{org141}
% ...
\And
\renewcommand\labelenumi{\textsuperscript{\theenumi}~}

\section*{Affiliation notes}
\renewcommand\theenumi{\roman{enumi}}
\begin{Authlist}
\item \Adef{org*}Deceased
\item \Adef{orgI}Dipartimento DET del Politecnico di Torino, Turin, Italy
\item \Adef{orgII}M.V. Lomonosov Moscow State University, D.V. Skobeltsyn Institute of Nuclear, Physics, Moscow, Russia
\item \Adef{orgIII}Department of Applied Physics, Aligarh Muslim University, Aligarh, India
\item \Adef{orgIV}Institute of Theoretical Physics, University of Wroclaw, Poland
\end{Authlist}

\section*{Collaboration Institutes}
\renewcommand\theenumi{\arabic{enumi}~}
\begin{Authlist}
\item \Idef{org1}A.I. Alikhanyan National Science Laboratory (Yerevan Physics Institute) Foundation, Yerevan, Armenia
\item \Idef{org2}Bogolyubov Institute for Theoretical Physics, National Academy of Sciences of Ukraine, Kiev, Ukraine
\item \Idef{org3}Bose Institute, Department of Physics  and Centre for Astroparticle Physics and Space Science (CAPSS), Kolkata, India
\item \Idef{org4}Budker Institute for Nuclear Physics, Novosibirsk, Russia
\item \Idef{org5}California Polytechnic State University, San Luis Obispo, California, United States
\item \Idef{org6}Central China Normal University, Wuhan, China
\item \Idef{org7}Centre de Calcul de l'IN2P3, Villeurbanne, Lyon, France
\item \Idef{org8}Centro de Aplicaciones Tecnol\'{o}gicas y Desarrollo Nuclear (CEADEN), Havana, Cuba
\item \Idef{org9}Centro de Investigaci\'{o}n y de Estudios Avanzados (CINVESTAV), Mexico City and M\'{e}rida, Mexico
\item \Idef{org10}Centro Fermi - Museo Storico della Fisica e Centro Studi e Ricerche ``Enrico Fermi', Rome, Italy
\item \Idef{org11}Chicago State University, Chicago, Illinois, United States
\item \Idef{org12}China Institute of Atomic Energy, Beijing, China
\item \Idef{org13}Chonbuk National University, Jeonju, Republic of Korea
\item \Idef{org14}Comenius University Bratislava, Faculty of Mathematics, Physics and Informatics, Bratislava, Slovakia
\item \Idef{org15}COMSATS University Islamabad, Islamabad, Pakistan
\item \Idef{org16}Creighton University, Omaha, Nebraska, United States
\item \Idef{org17}Department of Physics, Aligarh Muslim University, Aligarh, India
\item \Idef{org18}Department of Physics, Pusan National University, Pusan, Republic of Korea
\item \Idef{org19}Department of Physics, Sejong University, Seoul, Republic of Korea
\item \Idef{org20}Department of Physics, University of California, Berkeley, California, United States
\item \Idef{org21}Department of Physics, University of Oslo, Oslo, Norway
\item \Idef{org22}Department of Physics and Technology, University of Bergen, Bergen, Norway
\item \Idef{org23}Dipartimento di Fisica dell'Universit\`{a} 'La Sapienza' and Sezione INFN, Rome, Italy
\item \Idef{org24}Dipartimento di Fisica dell'Universit\`{a} and Sezione INFN, Cagliari, Italy
\item \Idef{org25}Dipartimento di Fisica dell'Universit\`{a} and Sezione INFN, Trieste, Italy
\item \Idef{org26}Dipartimento di Fisica dell'Universit\`{a} and Sezione INFN, Turin, Italy
\item \Idef{org27}Dipartimento di Fisica e Astronomia dell'Universit\`{a} and Sezione INFN, Bologna, Italy
\item \Idef{org28}Dipartimento di Fisica e Astronomia dell'Universit\`{a} and Sezione INFN, Catania, Italy
\item \Idef{org29}Dipartimento di Fisica e Astronomia dell'Universit\`{a} and Sezione INFN, Padova, Italy
\item \Idef{org30}Dipartimento di Fisica `E.R.~Caianiello' dell'Universit\`{a} and Gruppo Collegato INFN, Salerno, Italy
\item \Idef{org31}Dipartimento DISAT del Politecnico and Sezione INFN, Turin, Italy
\item \Idef{org32}Dipartimento di Scienze e Innovazione Tecnologica dell'Universit\`{a} del Piemonte Orientale and INFN Sezione di Torino, Alessandria, Italy
\item \Idef{org33}Dipartimento Interateneo di Fisica `M.~Merlin' and Sezione INFN, Bari, Italy
\item \Idef{org34}European Organization for Nuclear Research (CERN), Geneva, Switzerland
\item \Idef{org35}Faculty of Electrical Engineering, Mechanical Engineering and Naval Architecture, University of Split, Split, Croatia
\item \Idef{org36}Faculty of Engineering and Science, Western Norway University of Applied Sciences, Bergen, Norway
\item \Idef{org37}Faculty of Nuclear Sciences and Physical Engineering, Czech Technical University in Prague, Prague, Czech Republic
\item \Idef{org38}Faculty of Science, P.J.~\v{S}af\'{a}rik University, Ko\v{s}ice, Slovakia
\item \Idef{org39}Frankfurt Institute for Advanced Studies, Johann Wolfgang Goethe-Universit\"{a}t Frankfurt, Frankfurt, Germany
\item \Idef{org40}Gangneung-Wonju National University, Gangneung, Republic of Korea
\item \Idef{org41}Gauhati University, Department of Physics, Guwahati, India
\item \Idef{org42}Helmholtz-Institut f\"{u}r Strahlen- und Kernphysik, Rheinische Friedrich-Wilhelms-Universit\"{a}t Bonn, Bonn, Germany
\item \Idef{org43}Helsinki Institute of Physics (HIP), Helsinki, Finland
\item \Idef{org44}High Energy Physics Group,  Universidad Aut\'{o}noma de Puebla, Puebla, Mexico
\item \Idef{org45}Hiroshima University, Hiroshima, Japan
\item \Idef{org46}Hochschule Worms, Zentrum  f\"{u}r Technologietransfer und Telekommunikation (ZTT), Worms, Germany
\item \Idef{org47}Horia Hulubei National Institute of Physics and Nuclear Engineering, Bucharest, Romania
\item \Idef{org48}Indian Institute of Technology Bombay (IIT), Mumbai, India
\item \Idef{org49}Indian Institute of Technology Indore, Indore, India
\item \Idef{org50}Indonesian Institute of Sciences, Jakarta, Indonesia
\item \Idef{org51}INFN, Laboratori Nazionali di Frascati, Frascati, Italy
\item \Idef{org52}INFN, Sezione di Bari, Bari, Italy
\item \Idef{org53}INFN, Sezione di Bologna, Bologna, Italy
\item \Idef{org54}INFN, Sezione di Cagliari, Cagliari, Italy
\item \Idef{org55}INFN, Sezione di Catania, Catania, Italy
\item \Idef{org56}INFN, Sezione di Padova, Padova, Italy
\item \Idef{org57}INFN, Sezione di Roma, Rome, Italy
\item \Idef{org58}INFN, Sezione di Torino, Turin, Italy
\item \Idef{org59}INFN, Sezione di Trieste, Trieste, Italy
\item \Idef{org60}Inha University, Incheon, Republic of Korea
\item \Idef{org61}Institut de Physique Nucl\'{e}aire d'Orsay (IPNO), Institut National de Physique Nucl\'{e}aire et de Physique des Particules (IN2P3/CNRS), Universit\'{e} de Paris-Sud, Universit\'{e} Paris-Saclay, Orsay, France
\item \Idef{org62}Institute for Nuclear Research, Academy of Sciences, Moscow, Russia
\item \Idef{org63}Institute for Subatomic Physics, Utrecht University/Nikhef, Utrecht, Netherlands
\item \Idef{org64}Institute for Theoretical and Experimental Physics, Moscow, Russia
\item \Idef{org65}Institute of Experimental Physics, Slovak Academy of Sciences, Ko\v{s}ice, Slovakia
\item \Idef{org66}Institute of Physics, Homi Bhabha National Institute, Bhubaneswar, India
\item \Idef{org67}Institute of Physics of the Czech Academy of Sciences, Prague, Czech Republic
\item \Idef{org68}Institute of Space Science (ISS), Bucharest, Romania
\item \Idef{org69}Institut f\"{u}r Kernphysik, Johann Wolfgang Goethe-Universit\"{a}t Frankfurt, Frankfurt, Germany
\item \Idef{org70}Instituto de Ciencias Nucleares, Universidad Nacional Aut\'{o}noma de M\'{e}xico, Mexico City, Mexico
\item \Idef{org71}Instituto de F\'{i}sica, Universidade Federal do Rio Grande do Sul (UFRGS), Porto Alegre, Brazil
\item \Idef{org72}Instituto de F\'{\i}sica, Universidad Nacional Aut\'{o}noma de M\'{e}xico, Mexico City, Mexico
\item \Idef{org73}iThemba LABS, National Research Foundation, Somerset West, South Africa
\item \Idef{org74}Johann-Wolfgang-Goethe Universit\"{a}t Frankfurt Institut f\"{u}r Informatik, Fachbereich Informatik und Mathematik, Frankfurt, Germany
\item \Idef{org75}Joint Institute for Nuclear Research (JINR), Dubna, Russia
\item \Idef{org76}Korea Institute of Science and Technology Information, Daejeon, Republic of Korea
\item \Idef{org77}KTO Karatay University, Konya, Turkey
\item \Idef{org78}Laboratoire de Physique Subatomique et de Cosmologie, Universit\'{e} Grenoble-Alpes, CNRS-IN2P3, Grenoble, France
\item \Idef{org79}Lawrence Berkeley National Laboratory, Berkeley, California, United States
\item \Idef{org80}Lund University Department of Physics, Division of Particle Physics, Lund, Sweden
\item \Idef{org81}Nagasaki Institute of Applied Science, Nagasaki, Japan
\item \Idef{org82}Nara Women{'}s University (NWU), Nara, Japan
\item \Idef{org83}National and Kapodistrian University of Athens, School of Science, Department of Physics , Athens, Greece
\item \Idef{org84}National Centre for Nuclear Research, Warsaw, Poland
\item \Idef{org85}National Institute of Science Education and Research, Homi Bhabha National Institute, Jatni, India
\item \Idef{org86}National Nuclear Research Center, Baku, Azerbaijan
\item \Idef{org87}National Research Centre Kurchatov Institute, Moscow, Russia
\item \Idef{org88}Niels Bohr Institute, University of Copenhagen, Copenhagen, Denmark
\item \Idef{org89}Nikhef, National institute for subatomic physics, Amsterdam, Netherlands
\item \Idef{org90}NRC Kurchatov Institute IHEP, Protvino, Russia
\item \Idef{org91}NRNU Moscow Engineering Physics Institute, Moscow, Russia
\item \Idef{org92}Nuclear Physics Group, STFC Daresbury Laboratory, Daresbury, United Kingdom
\item \Idef{org93}Nuclear Physics Institute of the Czech Academy of Sciences, \v{R}e\v{z} u Prahy, Czech Republic
\item \Idef{org94}Oak Ridge National Laboratory, Oak Ridge, Tennessee, United States
\item \Idef{org95}Ohio State University, Columbus, Ohio, United States
\item \Idef{org96}Petersburg Nuclear Physics Institute, Gatchina, Russia
\item \Idef{org97}Physics department, Faculty of science, University of Zagreb, Zagreb, Croatia
\item \Idef{org98}Physics Department, Panjab University, Chandigarh, India
\item \Idef{org99}Physics Department, University of Jammu, Jammu, India
\item \Idef{org100}Physics Department, University of Rajasthan, Jaipur, India
\item \Idef{org101}Physikalisches Institut, Eberhard-Karls-Universit\"{a}t T\"{u}bingen, T\"{u}bingen, Germany
\item \Idef{org102}Physikalisches Institut, Ruprecht-Karls-Universit\"{a}t Heidelberg, Heidelberg, Germany
\item \Idef{org103}Physik Department, Technische Universit\"{a}t M\"{u}nchen, Munich, Germany
\item \Idef{org104}Politecnico di Bari, Bari, Italy
\item \Idef{org105}Research Division and ExtreMe Matter Institute EMMI, GSI Helmholtzzentrum f\"ur Schwerionenforschung GmbH, Darmstadt, Germany
\item \Idef{org106}Rudjer Bo\v{s}kovi\'{c} Institute, Zagreb, Croatia
\item \Idef{org107}Russian Federal Nuclear Center (VNIIEF), Sarov, Russia
\item \Idef{org108}Saha Institute of Nuclear Physics, Homi Bhabha National Institute, Kolkata, India
\item \Idef{org109}School of Physics and Astronomy, University of Birmingham, Birmingham, United Kingdom
\item \Idef{org110}Secci\'{o}n F\'{\i}sica, Departamento de Ciencias, Pontificia Universidad Cat\'{o}lica del Per\'{u}, Lima, Peru
\item \Idef{org111}Shanghai Institute of Applied Physics, Shanghai, China
\item \Idef{org112}St. Petersburg State University, St. Petersburg, Russia
\item \Idef{org113}Stefan Meyer Institut f\"{u}r Subatomare Physik (SMI), Vienna, Austria
\item \Idef{org114}SUBATECH, IMT Atlantique, Universit\'{e} de Nantes, CNRS-IN2P3, Nantes, France
\item \Idef{org115}Suranaree University of Technology, Nakhon Ratchasima, Thailand
\item \Idef{org116}Technical University of Ko\v{s}ice, Ko\v{s}ice, Slovakia
\item \Idef{org117}Technische Universit\"{a}t M\"{u}nchen, Excellence Cluster 'Universe', Munich, Germany
\item \Idef{org118}The Henryk Niewodniczanski Institute of Nuclear Physics, Polish Academy of Sciences, Cracow, Poland
\item \Idef{org119}The University of Texas at Austin, Austin, Texas, United States
\item \Idef{org120}Universidad Aut\'{o}noma de Sinaloa, Culiac\'{a}n, Mexico
\item \Idef{org121}Universidade de S\~{a}o Paulo (USP), S\~{a}o Paulo, Brazil
\item \Idef{org122}Universidade Estadual de Campinas (UNICAMP), Campinas, Brazil
\item \Idef{org123}Universidade Federal do ABC, Santo Andre, Brazil
\item \Idef{org124}University College of Southeast Norway, Tonsberg, Norway
\item \Idef{org125}University of Cape Town, Cape Town, South Africa
\item \Idef{org126}University of Houston, Houston, Texas, United States
\item \Idef{org127}University of Jyv\"{a}skyl\"{a}, Jyv\"{a}skyl\"{a}, Finland
\item \Idef{org128}University of Liverpool, Liverpool, United Kingdom
\item \Idef{org129}University of Science and Techonology of China, Hefei, China
\item \Idef{org130}University of Tennessee, Knoxville, Tennessee, United States
\item \Idef{org131}University of the Witwatersrand, Johannesburg, South Africa
\item \Idef{org132}University of Tokyo, Tokyo, Japan
\item \Idef{org133}University of Tsukuba, Tsukuba, Japan
\item \Idef{org134}Universit\'{e} Clermont Auvergne, CNRS/IN2P3, LPC, Clermont-Ferrand, France
\item \Idef{org135}Universit\'{e} de Lyon, Universit\'{e} Lyon 1, CNRS/IN2P3, IPN-Lyon, Villeurbanne, Lyon, France
\item \Idef{org136}Universit\'{e} de Strasbourg, CNRS, IPHC UMR 7178, F-67000 Strasbourg, France, Strasbourg, France
\item \Idef{org137}Universit\'{e} Paris-Saclay Centre d'Etudes de Saclay (CEA), IRFU, D\'{e}partment de Physique Nucl\'{e}aire (DPhN), Saclay, France
\item \Idef{org138}Universit\`{a} degli Studi di Foggia, Foggia, Italy
\item \Idef{org139}Universit\`{a} degli Studi di Pavia, Pavia, Italy
\item \Idef{org140}Universit\`{a} di Brescia, Brescia, Italy
\item \Idef{org141}Variable Energy Cyclotron Centre, Homi Bhabha National Institute, Kolkata, India
\item \Idef{org142}Warsaw University of Technology, Warsaw, Poland
\item \Idef{org143}Wayne State University, Detroit, Michigan, United States
\item \Idef{org144}Westf\"{a}lische Wilhelms-Universit\"{a}t M\"{u}nster, Institut f\"{u}r Kernphysik, M\"{u}nster, Germany
\item \Idef{org145}Wigner Research Centre for Physics, Hungarian Academy of Sciences, Budapest, Hungary
\item \Idef{org146}Yale University, New Haven, Connecticut, United States
\item \Idef{org147}Yonsei University, Seoul, Republic of Korea
\end{Authlist}
\endgroup
\end{document}